\documentclass{article}

\usepackage{arxiv}

\usepackage[utf8]{inputenc} 
\usepackage[T1]{fontenc}    
\usepackage{hyperref}       
\usepackage{url}            
\usepackage{booktabs}       
\usepackage{amsfonts}       
\usepackage{nicefrac}       
\usepackage{microtype}      
\usepackage{lipsum}
\usepackage{rotating}
\usepackage{subcaption}
\usepackage{lscape}
\usepackage{graphicx}
    \usepackage{listings}
\graphicspath{ {./images/} }

\title{Comprehensive evaluation of no-reference image quality assessment algorithms
on KADID-10k database}

\author{
 Domonkos Varga 
}

\begin{document}
\maketitle
\begin{abstract}
The main goal of objective image quality assessment is to devise computational,
mathematical models which are able to predict perceptual image quality consistently
with subjective evaluations. The evaluation of objective image quality assessment
algorithms is based on experiments conducted on publicly available benchmark
databases. In this study, our goal is to give a comprehensive evaluation
about no-reference image quality assessment algorithms, whose original
source codes are available online, using the recently
published KADID-10k database which is one of the largest available benchmark
databases. Specifically, average PLCC, SROCC, and KROCC are reported which were
measured over 100 random train-test splits. Furthermore, the database was divided into a train (appx.
80\% of images) and a test set (appx. 20\% of images) with respect to the reference images.
So no semantic content overlap was between these two sets.
Our evaluation results may be helpful to obtain a clear understanding about the status of state-of-the-art no-reference image quality assessment methods.
\end{abstract}

\keywords{no-reference image quality assessment}

\section{Introduction}
There are three main categories of objective image quality
assessment (IQA) algorithms. Namely, IQA problems can be classified based on
the availability of the reference, pristine images. In 
\textit{full-reference image quality assessment (FR-IQA)}, the reference, pristine
(distortion-free) image and the distorted image are given to estimate the
perceptual quality of the distorted image. In contrast, 
\textit{no-reference image quality assessment (NR-IQA)} algorithms solely rely on
the distorted images. Similarly, 
in \textit{reduced-reference image quality assessment (RR-IQA)} the reference image
is not available, but partial information about it is known.

The evaluation of objective IQA algorithms is based on experiments carried out
on publicly available image quality databases which consist of images labelled with
quality scores. These databases can be divided into two categories based on
the distortion types. The first one contains dozens of pristine (distortion free),
reference images and the distorted images are derived from the reference images
using different levels of artificial distortions and different types of artificial
distortions, such as Gaussian blurring, JPEG compression, JPEG2000 compression,
color diffusion, \textit{etc.} The second one consists of authentically
distorted images captured by various imaging devices, such as mobile camera. As a
consequence, the images are afflicted by a highly complex mixture of
multiple distortions. Table \ref{table:iqadatabase} summarizes the main
characteristics of major publicly available image quality databases.

\begin{table*}[ht]
\caption{Comparison of several publicly available IQA databases.
} 
\centering 
\begin{center}
    \begin{tabular}{ |c|c|c|c|c|c|}
    \hline
Database&Ref. images&Test images&Resolution& Distortion levels&Number of distortions\\
    \hline
LIVE \cite{sheikh2006statistical}&29&779&$768\times512$&4-5&5 \\
A57 \cite{chandler2007vsnr}&3&54&$512\times512$&6&3 \\
Toyoma-MICT \cite{horita2011mict}&14&168&$768\times512$&6&2 \\
TID2008 \cite{ponomarenko2009tid2008}&25&1,700&$512\times384$&4&17 \\
CSIQ \cite{larson2010most} & 30& 866&$512\times512$&4-5 &6 \\
VCL-FER \cite{zaric2012vcl} & 23 & 552 &$683\times512$ &6& 4 \\
LIVE Multiple Distorted \cite{jayaraman2012objective}&15 &405 &$1280\times720$&3&2 \\
TID2013 \cite{ponomarenko2015image}&25&3,000&$512\times384$&5&24 \\
CID:IQ \cite{liu2014cid}&23&690&$800\times800$&5&6 \\
LIVE In the Wild \cite{ghadiyaram2015massive}&-&1,169&$500\times500$&-&N/A \\
MDID \cite{sun2017mdid}&20&1,600&$512\times384$&4&5 \\
KonIQ-10k \cite{lin2018koniq}&-&10,073&$1024\times768$&-&N/A \\
KADID-10k \cite{lin2019kadid} & 81 &10,125&$512\times384$&5&25\\
 \hline
 \end{tabular}
\end{center}
\label{table:iqadatabase}
\end{table*}

\subsection{Contributions}
\begin{figure}
    \centering
    \begin{subfigure}[b]{0.25\textwidth}
        \includegraphics[width=\textwidth]{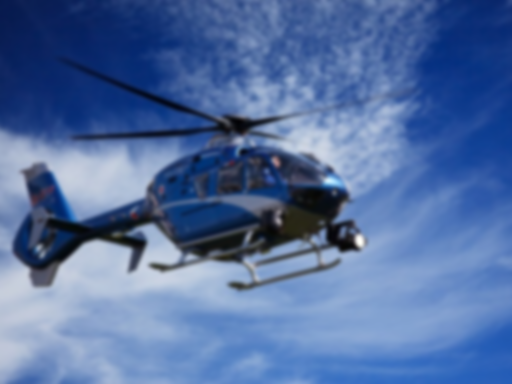}
        \caption{Gaussian blur.}
    \end{subfigure}
    ~ 
    \begin{subfigure}[b]{0.25\textwidth}
        \includegraphics[width=\textwidth]{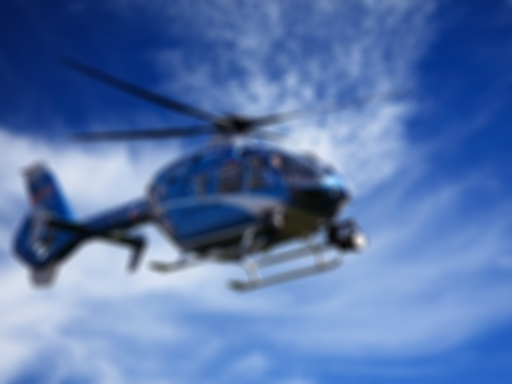}
        \caption{Lens blur.}
    \end{subfigure}
     ~ 
    \begin{subfigure}[b]{0.25\textwidth}
        \includegraphics[width=\textwidth]{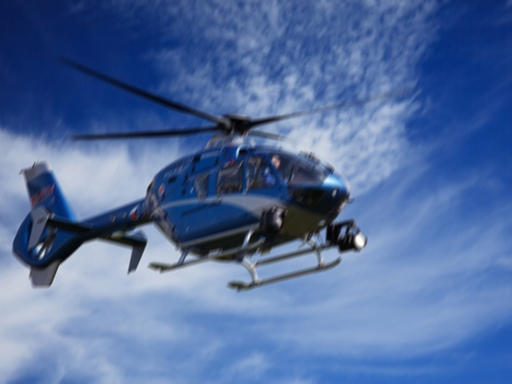}
        \caption{Motion blur.}
    \end{subfigure}
    
    \quad
    
    \begin{subfigure}[b]{0.25\textwidth}
        \includegraphics[width=\textwidth]{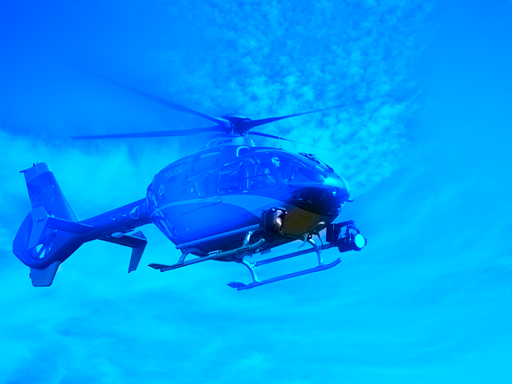}
        \caption{Color diffusion.}
    \end{subfigure}
    ~
    \begin{subfigure}[b]{0.25\textwidth}
        \includegraphics[width=\textwidth]{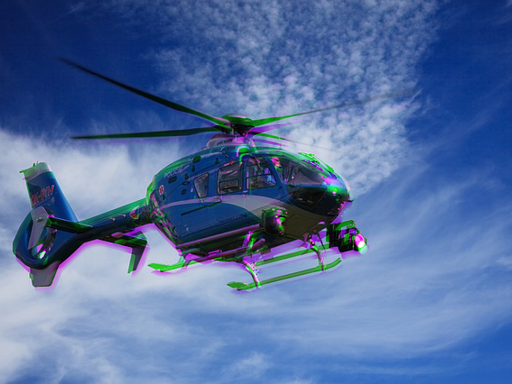}
        \caption{Color shift.}
    \end{subfigure}
     ~ 
    \begin{subfigure}[b]{0.25\textwidth}
        \includegraphics[width=\textwidth]{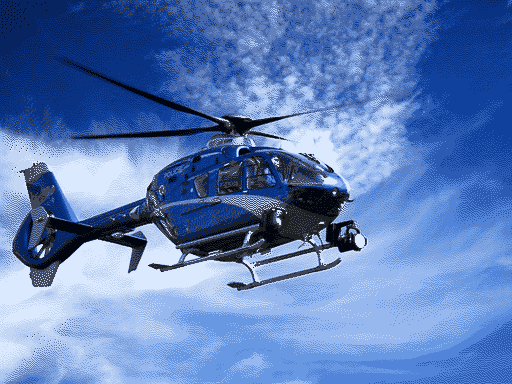}
        \caption{Color quantization.}
    \end{subfigure}
    
    \quad

    \begin{subfigure}[b]{0.25\textwidth}
        \includegraphics[width=\textwidth]{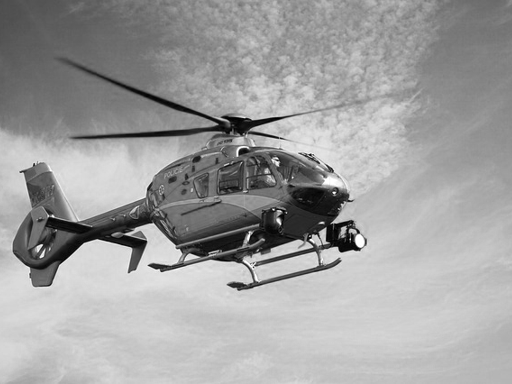}
        \caption{Color saturation 1.}
    \end{subfigure}
    ~
    \begin{subfigure}[b]{0.25\textwidth}
        \includegraphics[width=\textwidth]{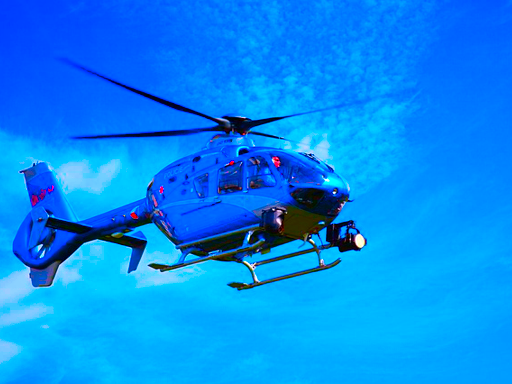}
        \caption{Color saturation 2.}
    \end{subfigure}
     ~ 
    \begin{subfigure}[b]{0.25\textwidth}
        \includegraphics[width=\textwidth]{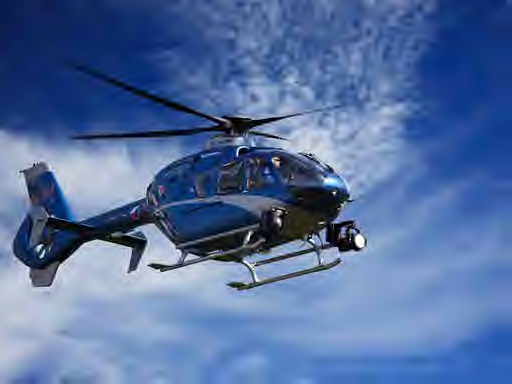}
        \caption{JPEG2000.}
    \end{subfigure}
    
    \quad
    
    \begin{subfigure}[b]{0.25\textwidth}
        \includegraphics[width=\textwidth]{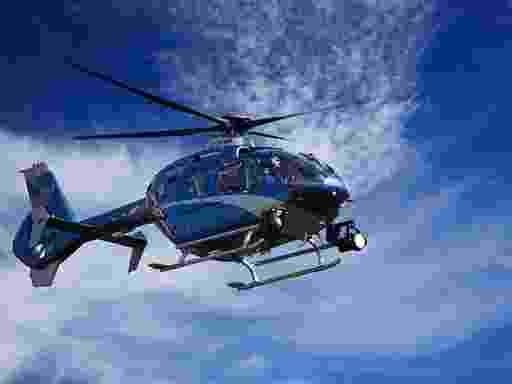}
        \caption{JPEG.}
    \end{subfigure}
    ~
    \begin{subfigure}[b]{0.25\textwidth}
        \includegraphics[width=\textwidth]{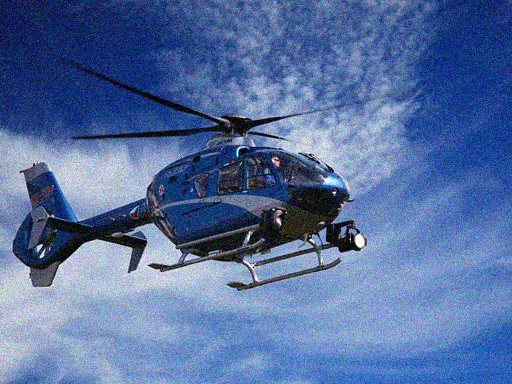}
        \caption{White noise.}
    \end{subfigure}
     ~ 
    \begin{subfigure}[b]{0.25\textwidth}
        \includegraphics[width=\textwidth]{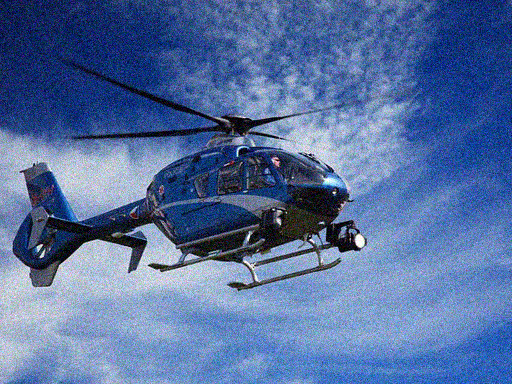}
        \caption{White noise color component.}
    \end{subfigure}
    
    \quad
    
    \begin{subfigure}[b]{0.25\textwidth}
        \includegraphics[width=\textwidth]{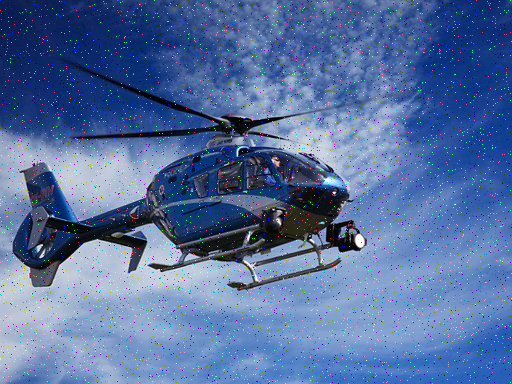}
        \caption{Impulse noise.}
    \end{subfigure}
    ~
    \begin{subfigure}[b]{0.25\textwidth}
        \includegraphics[width=\textwidth]{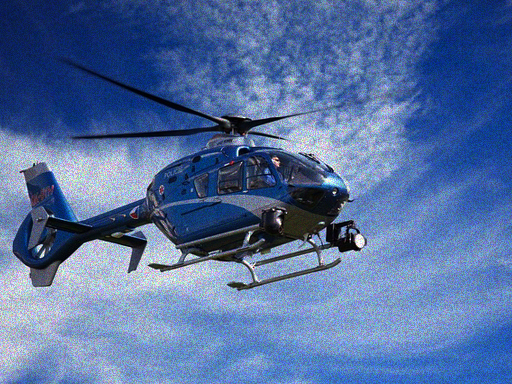}
        \caption{Multiplicative noise.}
    \end{subfigure}
     ~ 
    \begin{subfigure}[b]{0.25\textwidth}
        \includegraphics[width=\textwidth]{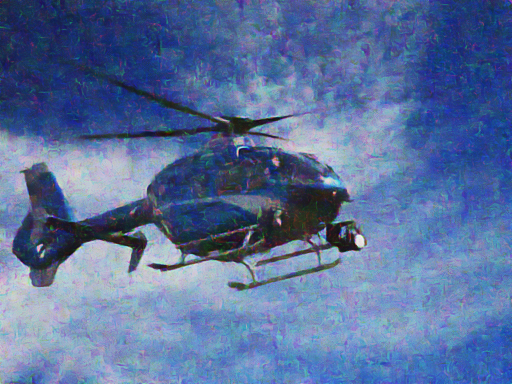}
        \caption{Denoise.}
    \end{subfigure}
    
    \caption{Distortion types of KADID-10k.}
    \label{fig:dist_1}
\end{figure}

\begin{figure}
    \centering
    \begin{subfigure}[b]{0.25\textwidth}
        \includegraphics[width=\textwidth]{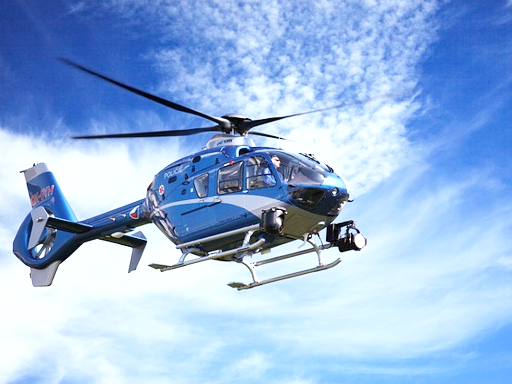}
        \caption{Brighten.}
    \end{subfigure}
    ~ 
    \begin{subfigure}[b]{0.25\textwidth}
        \includegraphics[width=\textwidth]{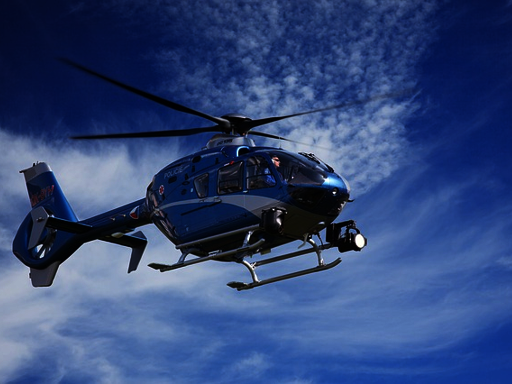}
        \caption{Darken.}
    \end{subfigure}
     ~ 
    \begin{subfigure}[b]{0.25\textwidth}
        \includegraphics[width=\textwidth]{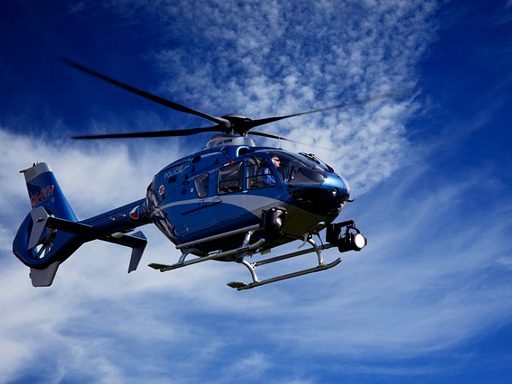}
        \caption{Mean shift.}
    \end{subfigure}
    
    \quad
    
    \begin{subfigure}[b]{0.25\textwidth}
        \includegraphics[width=\textwidth]{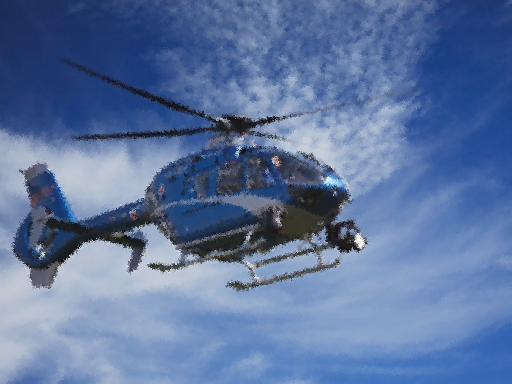}
        \caption{Jitter.}
    \end{subfigure}
    ~
    \begin{subfigure}[b]{0.25\textwidth}
        \includegraphics[width=\textwidth]{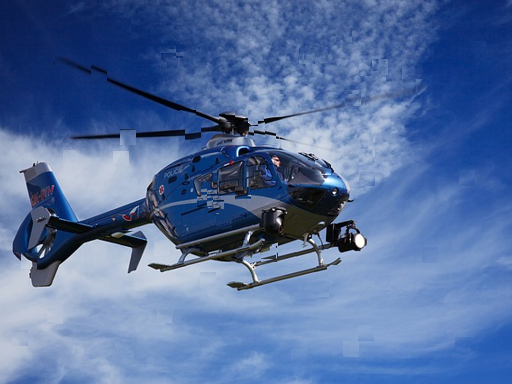}
        \caption{Non-eccentricity patch.}
    \end{subfigure}
     ~ 
    \begin{subfigure}[b]{0.25\textwidth}
        \includegraphics[width=\textwidth]{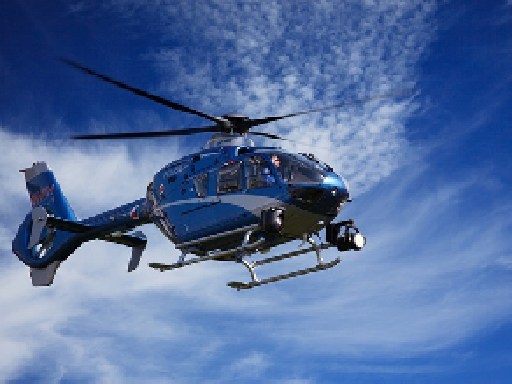}
        \caption{Pixelate.}
    \end{subfigure}
    
    \quad

    \begin{subfigure}[b]{0.25\textwidth}
        \includegraphics[width=\textwidth]{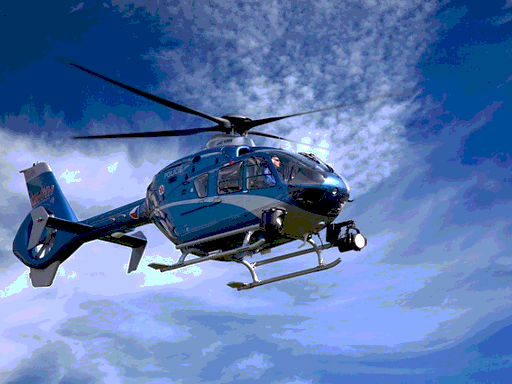}
        \caption{Quantization.}
    \end{subfigure}
    ~
    \begin{subfigure}[b]{0.25\textwidth}
        \includegraphics[width=\textwidth]{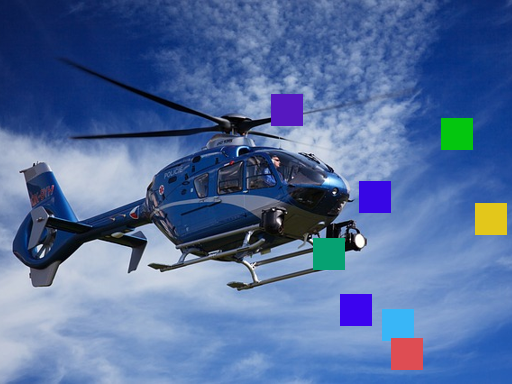}
        \caption{Color block.}
    \end{subfigure}
     ~ 
    \begin{subfigure}[b]{0.25\textwidth}
        \includegraphics[width=\textwidth]{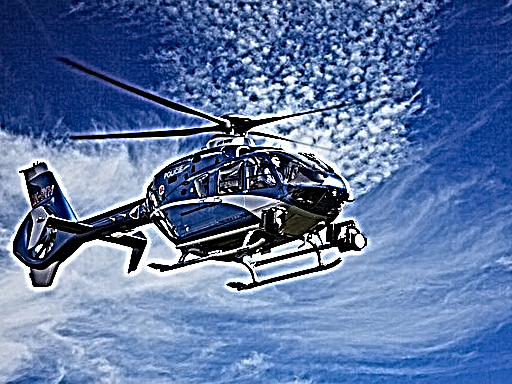}
        \caption{High sharpen.}
    \end{subfigure}
    
    \quad
    
    \begin{subfigure}[b]{0.25\textwidth}
        \includegraphics[width=\textwidth]{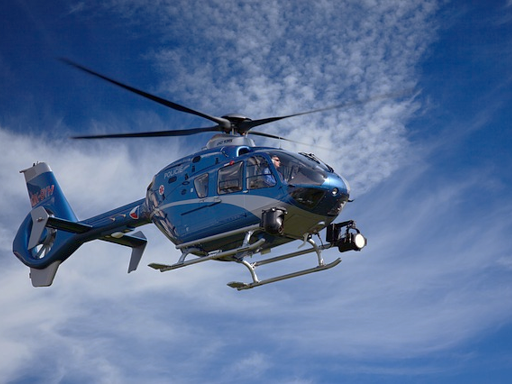}
        \caption{Contrast change.}
    \end{subfigure}
    
    \caption{Distortion types of KADID-10k.}
    \label{fig:dist_2}
\end{figure}

\begin{figure}
    \centering
    \begin{subfigure}[b]{0.25\textwidth}
        \includegraphics[width=\textwidth]{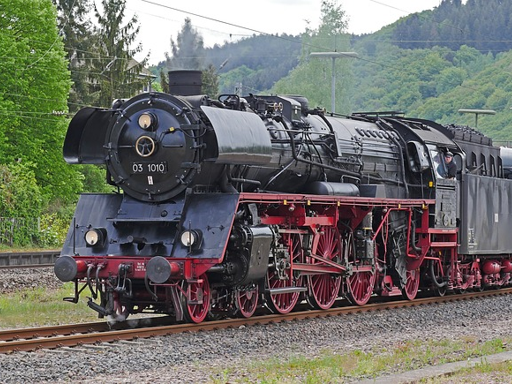}
        \caption{Reference image.}
    \end{subfigure}
    ~ 
    \begin{subfigure}[b]{0.25\textwidth}
        \includegraphics[width=\textwidth]{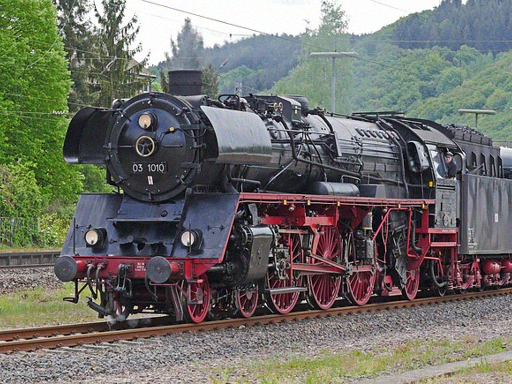}
        \caption{Level 1.}
    \end{subfigure}
     ~ 
    \begin{subfigure}[b]{0.25\textwidth}
        \includegraphics[width=\textwidth]{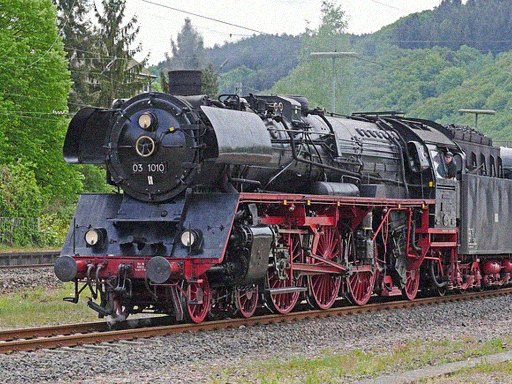}
        \caption{Level 2.}
    \end{subfigure}
    
    \quad
    
    \begin{subfigure}[b]{0.25\textwidth}
        \includegraphics[width=\textwidth]{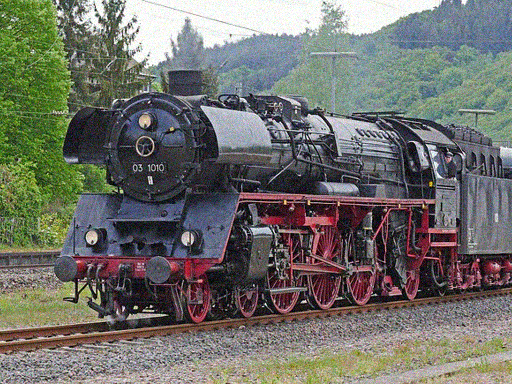}
        \caption{Level 3.}
    \end{subfigure}
    ~
    \begin{subfigure}[b]{0.25\textwidth}
        \includegraphics[width=\textwidth]{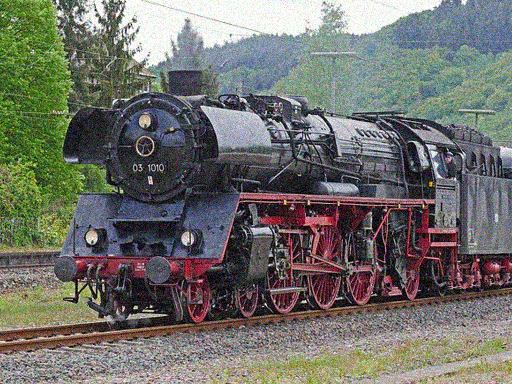}
        \caption{Level 4.}
    \end{subfigure}
    ~
    \begin{subfigure}[b]{0.25\textwidth}
        \includegraphics[width=\textwidth]{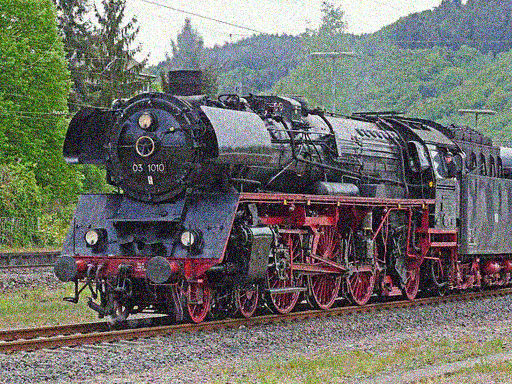}
        \caption{Level 5.}
    \end{subfigure}
    \caption{Illustration of distortion levels (white noise in color component).}
    \label{fig:levels}
\end{figure}

The goal of this study to provide a comprehensive evaluation of several
NR-IQA algorithms, including DIIVINE \cite{moorthy2011blind},
BLIINDS-II \cite{saad2012blind}, 
BRISQUE \cite{mittal2012no}, NIQE \cite{mittal2012making},
CurveletQA \cite{liu2014no},
SSEQ \cite{liu2014noS}, GRAD-LOG-CP \cite{xue2014blind}, PIQE \cite{venkatanath2015blind},
IL-NIQE \cite{zhang2015feature}, BMPRI \cite{min2018blind}, SPF-IQA \cite{varga2020no},
SCORER \cite{oszust2019local}, ENIQA \cite{chen2019no}, and
MultiGAP \cite{varga2020multi},
on the recently published
KADID-10k \cite{lin2019kadid} database. As one can see from
Table \ref{table:iqadatabase}, KADID-10k \cite{lin2019kadid} contains 81 reference
images and 10,125 distorted images using 25 different distortions in 5 levels.
Table \ref{table:distortions} summarizes the different distortion types
found in KADID-10k \cite{lin2019kadid}. Figure \ref{fig:dist_1} and \ref{fig:dist_2}
illustrate the distortion types of KADID-10k \cite{lin2019kadid}.
Figure \ref{fig:levels} depicts an illustration about the
five different distortion levels.

\begin{table*}[ht]
\caption{Distortion types found in KADID-10k \cite{lin2019kadid}.
} 
\centering 
\begin{center}
    \begin{tabular}{ |c|c|}
    \hline
Code&Distortion type\\
    \hline
1 & Gaussian blur \\
2 & Lens blur \\
3 & Motion blur \\
4 & Color diffusion \\
5 & Color shift \\
6 & Color quantization \\
7 & Color saturation 1. \\
8 & Color saturation 2. \\
9 & JPEG2000 \\
10 & JPEG \\
11 & White noise \\
12 & White noise in color component \\
13 & Impulse noise \\
14 & Multiplicative noise \\
15 & Denoise \\
16 & Brighten \\
17 & Darken \\
18 & Mean shift \\
19 & Jitter \\
20 & Non-eccentricity patch \\
21 & Pixelate \\
22 & Quantization \\
23 & Color block \\
24 & High sharpen \\
25 & Contrast change \\
 \hline
 \end{tabular}
\end{center}
\label{table:distortions}
\end{table*}

\subsection{Structure}
The rest of this study is organized as follows. After this short introduction,
Section \ref{sec:methods} gives a brief overview about NR-IQA methods with
a special attention to those algorithms which are evaluated in this study on
KADID-10k \cite{lin2019kadid} database. Section \label{ref:experimental}
demonstrates experimental results and analysis. Finally, a conclusion is drawn
in Section \ref{sec:conclusion}.
\section{Methods}
\label{sec:methods}
As already mentioned, the aim of NR-IQA is to estimate
the perceptual quality of a given image without any information about the pristine (distortion free),
reference image. Due to the lack of any knowledge about the reference medium,
NR-IQA is considered more challenging than FR-IQA or RR-IQA.
Although the reference image
is not available in NR-IQA, assumptions can be made about the distortion types found
in an
image, such as JPEG2000 compression noise. \textit{Distortion-specific} methods assume
one specific distortion type in the image, while \textit{general-purpose} algorithms
work over various types of distortions. Furthermore, general-purpose can be divided
into natural scene statistics (NSS) based, learning based, and human visual system (HVS)
based groups. Another classification of NR-IQA methods divides
existing methods into \textit{opinion-aware}
and \textit{opinion-unaware} classes. Opinion-aware methods utilize subject scores during
the training process, while opinion-unaware methods derive features from the
pristine, reference images of the database and perceptual quality of distorted images is quantified
as the deviation from the pristine images' features. In the followings, the examined methods
are briefly discussed.

\begin{figure}
\centering
\includegraphics[width=0.65\textwidth]{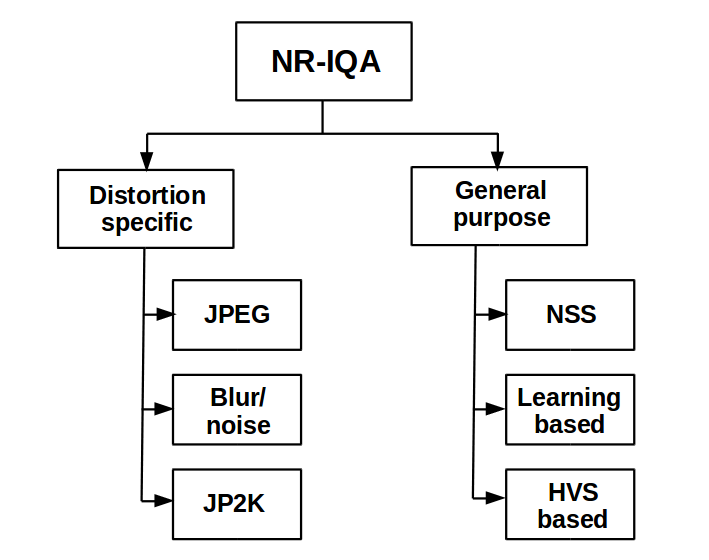}
\caption{Classification of no-reference image quality assessment.}
\label{fig:classification}
\end{figure}

\textbf{DIIVINE} \cite{moorthy2011blind} is a two-stage framework
which involves distortion
identification and distortion-specific quality assessment. It is
based on NSS. Namely, a set of
neighboring wavelet coefficients were modelled by a Gaussian scale mixture model.
Moreover, steerable pyramid decomposition was used to extract
statistics from the distorted
images.

\textbf{BLIINDS-II} \cite{saad2012blind} derives NSS features by discrete cosine transform
coefficients modeling using generalized Gaussian distribution. The parameters of the
generalized Gaussian distribution were applied as quality-aware features.

\textbf{BRISQUE} \cite{mittal2012no} applies scene statistics of locally
normalized luminance coefficients to
train a support vector regressor (SVR) for perceptual quality prediction.

\textbf{NIQE} \cite{mittal2012making} measures the distance between the 
natural scene statistics (NSS) based features
calculated from the pristine images to the features extracted from the input image. The
features are modeled as multi-dimensional Gaussian distributions.

\textbf{CurveletQA} \cite{liu2014no} extracted statistical
features (the coordinates of the maxima of the log-histograms of the curvelet coefficients,
the energy distributions of both orientation and scale in the curvelet domain) from 
the image's curvelet
representation. Moreover, image distortion and quality prediction stages are trained
using a support vector machine (SVM).

\textbf{SSEQ} \cite{liu2014noS} contains an image distortion and quality prediction engine.
Furthermore, it extracts a 12-dimensional local entropy feature vector.

\textbf{GRAD-LOG-CP} \cite{xue2014blind} utilizes the joint statistics of gradient magnitude
map and the Laplacian of Gaussian features to train a support vector regressor (SVR) for
perceptual image quality prediction.

\textbf{PIQE} \cite{venkatanath2015blind} is an opinion-unaware method and calculates perceptual
quality of an image through block-wise distortion estimation. First, the mean
subtracted contrast normalized (MSCN) coefficients are determined for each pixel
in the input image. Second, the input image is
divided into $16\times16$ blocks and high spatially
active blocks are identified relying on the MSCN coefficients.
In each block, distortion is evaluated
due to blocking artifacts and noise relying on the
MSCN coefficients. A threshold criteria is also
applied to classify blocks as distorted
(blocking artifacts, Gaussian noise) blocks. The
quality score is computed as the mean of scores in the distorted blocks.

\textbf{IL-NIQE} \cite{zhang2015feature} is an opinion-unaware method.
It integrated natural image statistics features from multiple sources,
such as normalized luminance, mean subtracted and contrast normalized
coefficients, gradient statistics, statistics of log-Gabor filter
responses, and statistics of colors. Subsequently, a multivariate Gaussian
model is learned from pristine image patches. Perceptual quality
is quantified by measuring the deviation from the learned distribution
using a Bhattacharyya-like distance.

\textbf{BMPRI} \cite{min2018blind} introduced the concept of multiple pseudo reference
images (MPRI). Specifically, the distorted images were further degraded. Subsequently,
similarities between the distorted image and the MPRIs were measured.
To this end, a traditional
FR-IQA metric was applied. Specifically, local binary pattern features
were computed to describe the similarities between the distorted image and the MPRIs.
Finally, the similarity scores
were aggregated to obtain the input image's perceptual quality.

\textbf{SPF-IQA} \cite{varga2020no} extracted different
statistical (fractal dimension distribution, first digit distribution
in gradient magnitude domain, first digit distribution in wavelet domain, color
statistics) and
perceptual features (colorfulness, global contrast factor,
dark channel feature, entropy, mean of phase congruency)
from the input image and fused them together. Finally, the
fused feature vector is mapped onto perceptual quality scores with the help
of Gaussian process regression (GPR) using rational quadratic kernel
function.

\textbf{SCORER} \cite{oszust2019local} proposed a set of derivative kernels
which were utilized to filter the \textit{Y}, \textit{Cb}, and
\textit{Cr} channels of the input image. As a result, a set of filtered
images was obtained for further processing. Subsequently, \textit{N}
interest points were detected on each filtered image relying on 
features from accelerated segment test (FAST) \cite{rosten2006machine}. Specifically,
each interest point is used to describe a $3\times3$ block
around the interest point
by taking all values from the block.
The extracted $2400$-dimensional feature vectors are mapped onto
perceptual quality scores with the help of a trained support vector regressor (SVR).

\textbf{ENIQA} \cite{chen2019no} extracts features 
in two different domains. Namely, mutual information
between color channels and the two-dimensional entropy is determined first.
Subsequently, two-dimensional entropy and the mutual information
of the filtered sub-band images are determined. Based on the extracted
features, a support vector machine (SVM) and a support vector regressor (SVR)
is trained for distortion and quality prediction, respectively.

In \textbf{MultiGAP} \cite{varga2020multi}, an input image is run through an
Inception-V3 pretrained convolutional neural network which carries
out all its defined operations. Moreover, global average pooling layers are attached
to each Inception module to extract image resolution independent features.
Subsequently, the features of the Inception modules are concatenated and mapped
onto perceptual quality scores with the help of an
support vector regressor (SVR)
with Gaussian kernel
function. In this study, results obtained by Gaussain process regression (GPR)
head with rational quadratic function is also presented.

\section{Experimental results and analysis}
\label{sec:experimental}
In this section, we report on the experimental results obtained on KADID-10k \cite{lin2019kadid}
database. The rest of this section is organized as follows.
In Subsection \ref{sec:eval}, the evaluation metrics are defined. 
Subsection \ref{sec:exp} describes the experimental setup. 
In Subsection \ref{sec:overall},
the performance of the examined NR-IQA algorithms is reported on the entire
KADID-10k \cite{lin2019kadid}. Subsection \ref{sec:level} reports the
performance
on individual distortion levels, while
Subsection \ref{sec:distortion} presents the experimental
results with respect to the distortion types.
\subsection{Evaluation metrics}
\label{sec:eval}
As already mentioned, the evaluation of IQA algorithms is based
on the correlation between the predicted and ground-truth scores measured
on an image quality database.
In the literature, there are three major evaluation: Pearson's linear correlation
coefficient (PLCC), Spearman's rank order correlation coefficient (SROCC), and
Kendall's rank order correlation coefficient (KROCC). The latter two measure the
prediction monotonicity of an IQA method, because they operate merely on the rank of the
data points and ignore the relative distance between data points. MATLAB provides
functions for the computation of these performance metrics.
\begin{lstlisting}[language=Matlab]
PLCC = corr(groundScores, predictedScores);
SROCC= corr(groundScores, predictedScores, 'Type', 'Spearman');
KROCC= corr(groundScores, predictedScores, 'Type', 'Kendall');
\end{lstlisting}

The PLCC between two data sets, $A$ and $B$, is defined as
\begin{equation}
PLCC(A,B)=\frac{\sum_{i=1}^n (A_i - \bar{A}) (B_i - \bar{B})}{\sqrt{\sum_{i=1}^n (A_i - \bar{A})^2} \sqrt{\sum_{i=1}^n (B_i - \bar{B})^2}},    
\end{equation}
where $\bar{A}=\frac{1}{n}\sum_{i=1}^n A_i$ and
$\bar{B}=\frac{1}{n}\sum_{i=1}^n B_i$. SROCC between $A$ and $B$
datasets is defined as
\begin{equation}
SROCC(A,B)=PLCC(rank(A), rank(B)),    
\end{equation}
where the $rank(\cdot)$ function gives back a vector whose $i$th element
is the rank of the $i$th element in the input vector.
KROCC is defined as
\begin{equation}
KROCC(A,B)=\frac{n_c-n_d}{\frac{1}{2}n(n-1)},    
\end{equation}
where $n$ is the length of $A$ and $B$, $n_c$ denotes the number
of concordant pairs between $A$ and $B$, and $n_d$ is the number
of discordant pairs.
\subsection{Experimental setup}
\label{sec:exp}
For learning-based methods, the distorted images of
KADID-10k \cite{lin2019kadid} were divided into a training (appx. 80\% of images)
and a test set (appx. 20\% of images) with respect to the reference images.
As a consequence, there was no semantic overlap between these two sets. Moreover,
average PLCC, SROCC, and KROCC were measured over 100 random train-test splits.
In contrast, opinion-unaware methods were trained on the reference images and
tested on the distorted images. Furthermore, PLCC, SROCC, and KROCC are reported.
\subsection{Overall performance}
\label{sec:overall}
\begin{figure}
    \centering
    \begin{subfigure}[b]{0.45\textwidth}
        \includegraphics[width=\textwidth]{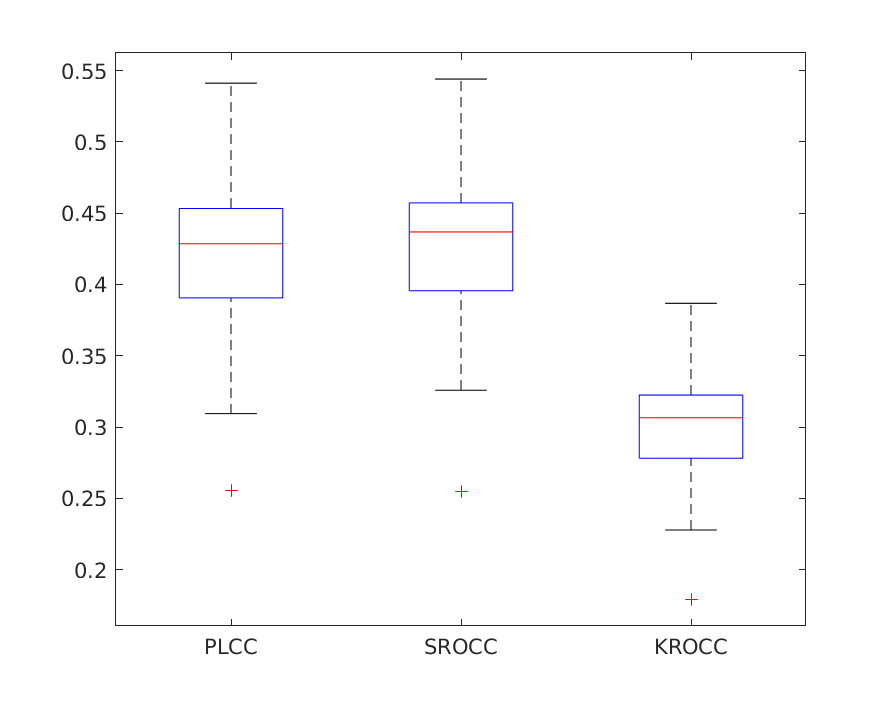}
        \caption{DIIVINE.}
    \end{subfigure}
    ~ 
    \begin{subfigure}[b]{0.45\textwidth}
        \includegraphics[width=\textwidth]{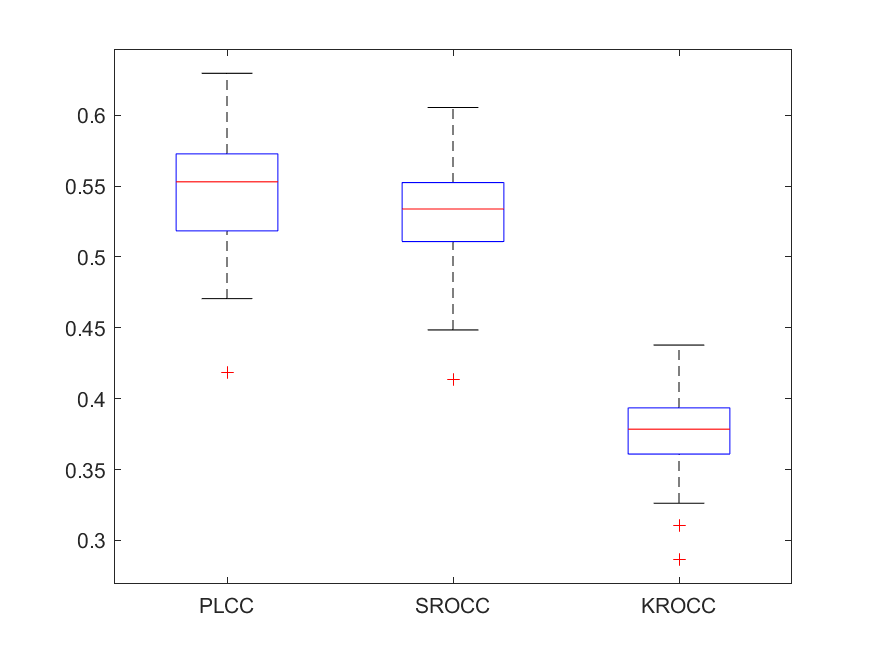}
        \caption{BLIINDS-II.}
    \end{subfigure}

    \quad
    
    \begin{subfigure}[b]{0.45\textwidth}
        \includegraphics[width=\textwidth]{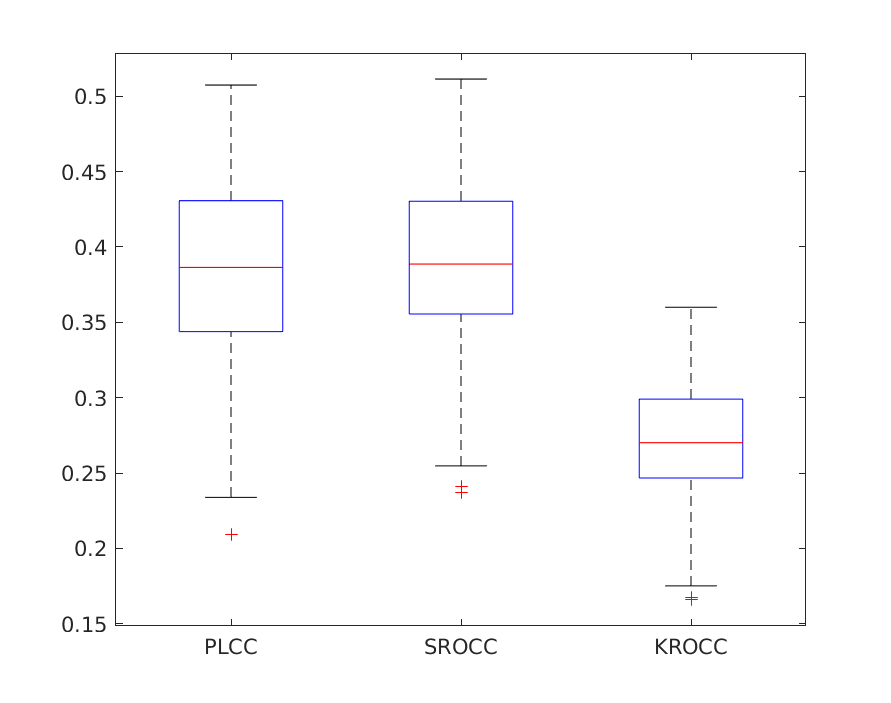}
        \caption{BRISQUE.}
    \end{subfigure}
    ~
    \begin{subfigure}[b]{0.45\textwidth}
        \includegraphics[width=\textwidth]{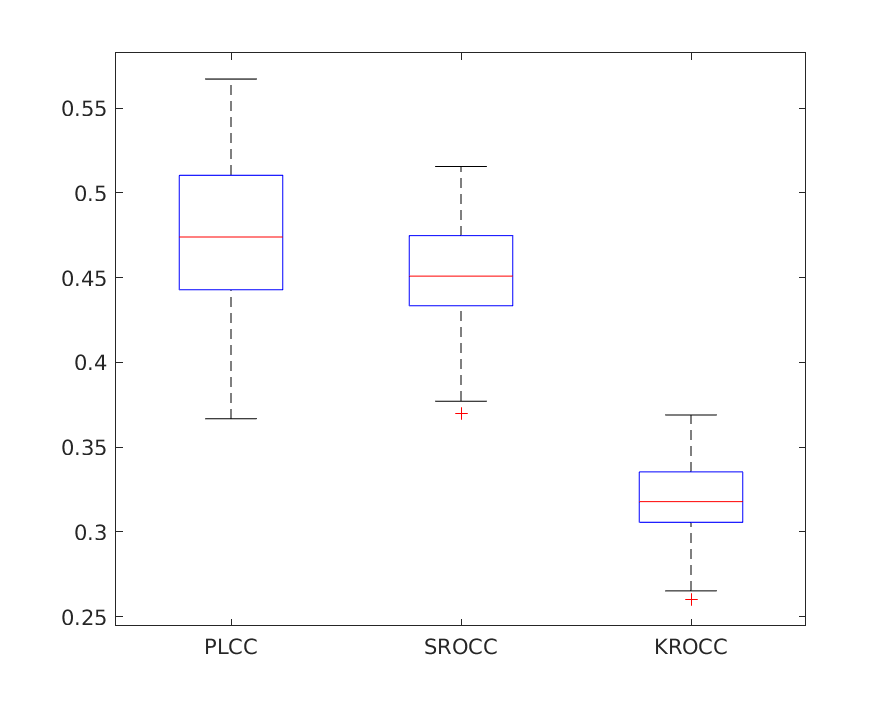}
        \caption{CurveletQA.}
    \end{subfigure}

    \quad

    \begin{subfigure}[b]{0.45\textwidth}
        \includegraphics[width=\textwidth]{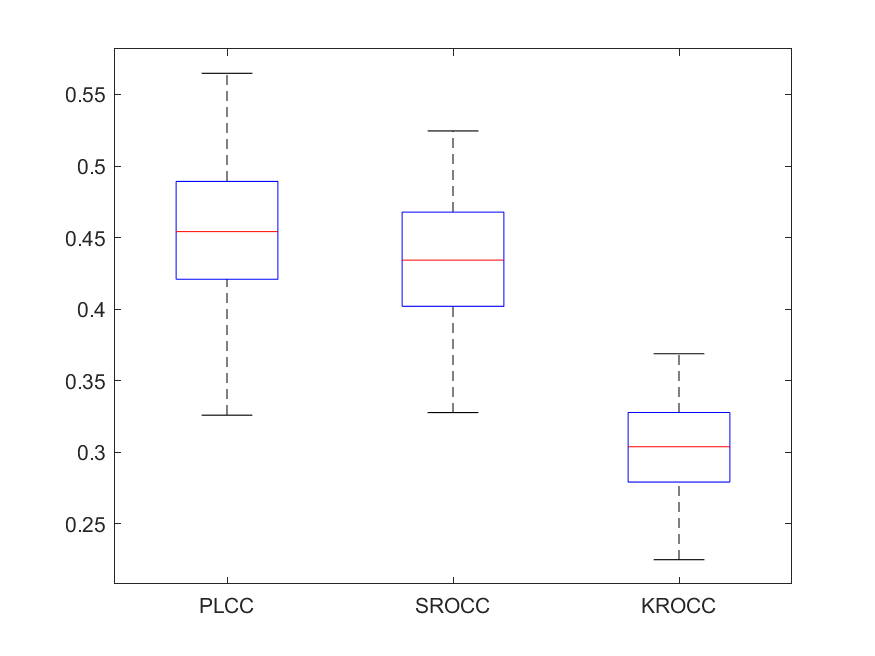}
        \caption{SSEQ.}
    \end{subfigure}
    ~
    \begin{subfigure}[b]{0.45\textwidth}
        \includegraphics[width=\textwidth]{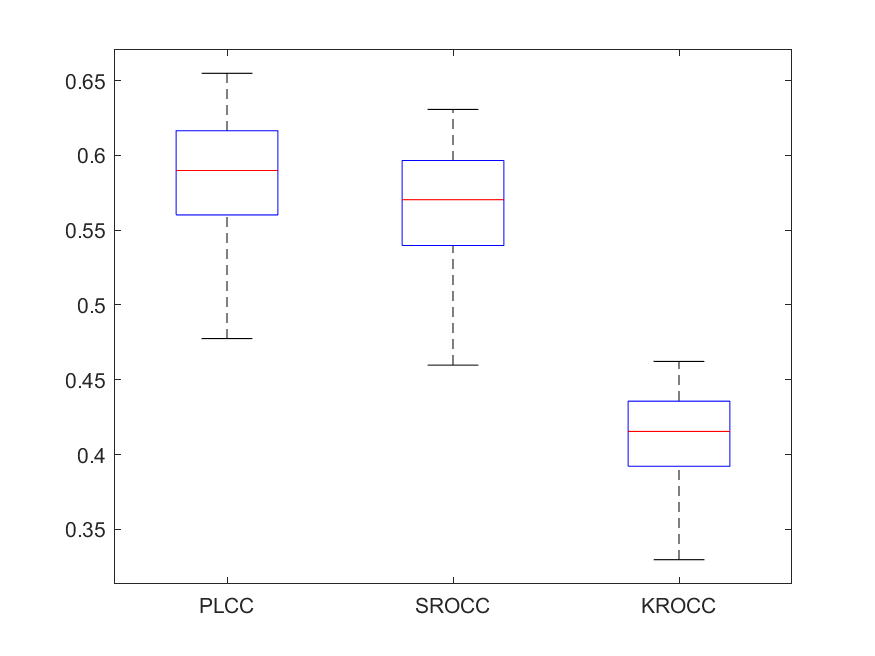}
        \caption{GRAD-LOG-CP.}
    \end{subfigure}
    \caption{Box plots of the measured PLCC, SROCC, and KROCC values. 100 random train-test splits.}
    \label{fig:boxplot1}
\end{figure}

\begin{figure}
    \centering
    \begin{subfigure}[b]{0.45\textwidth}
        \includegraphics[width=\textwidth]{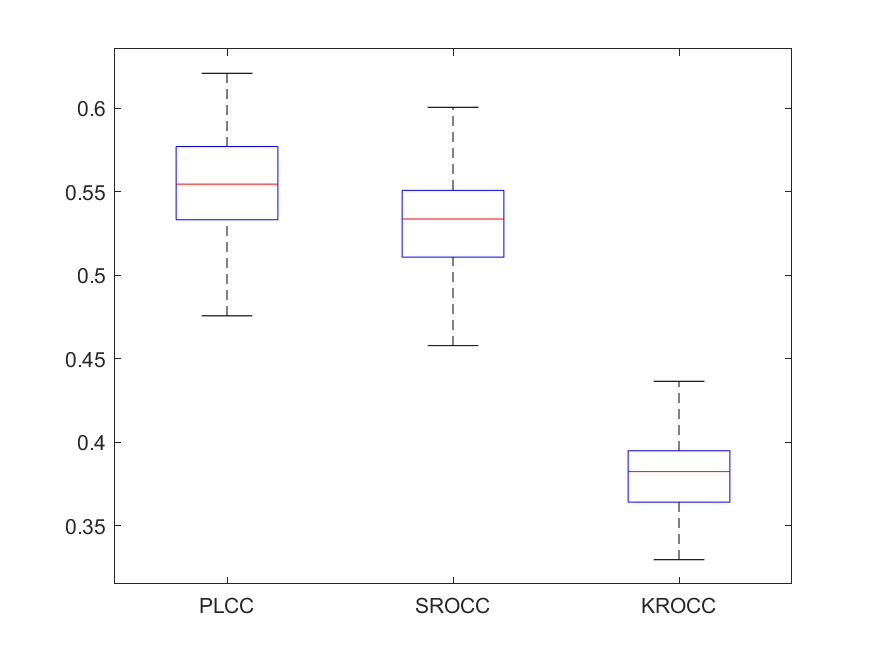}
        \caption{BMPRI.}
    \end{subfigure}
    ~ 
    \begin{subfigure}[b]{0.45\textwidth}
        \includegraphics[width=\textwidth]{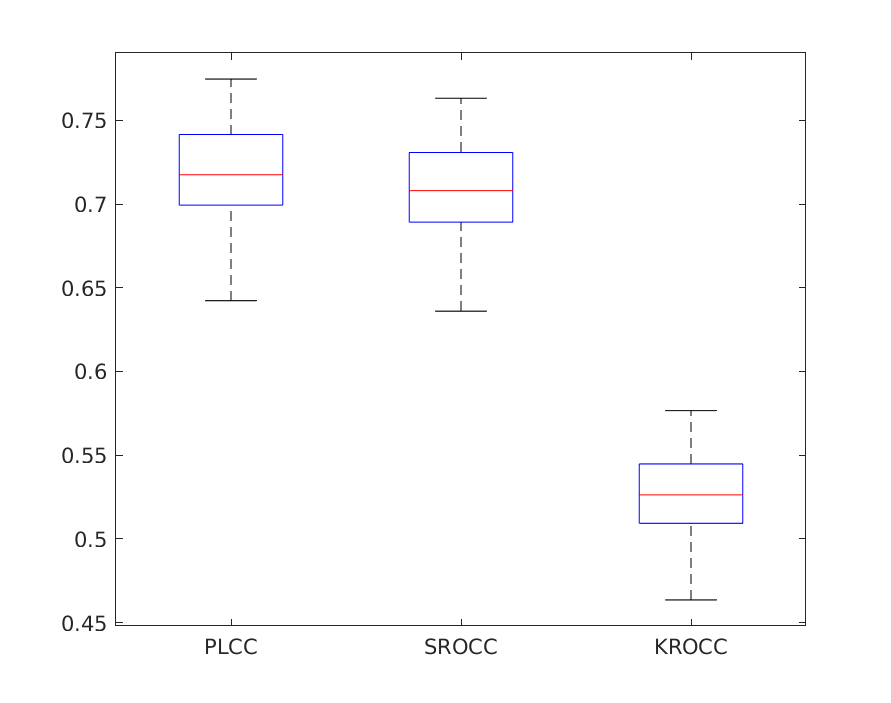}
        \caption{SPF-IQA.}
    \end{subfigure}

    \quad
    
    \begin{subfigure}[b]{0.45\textwidth}
        \includegraphics[width=\textwidth]{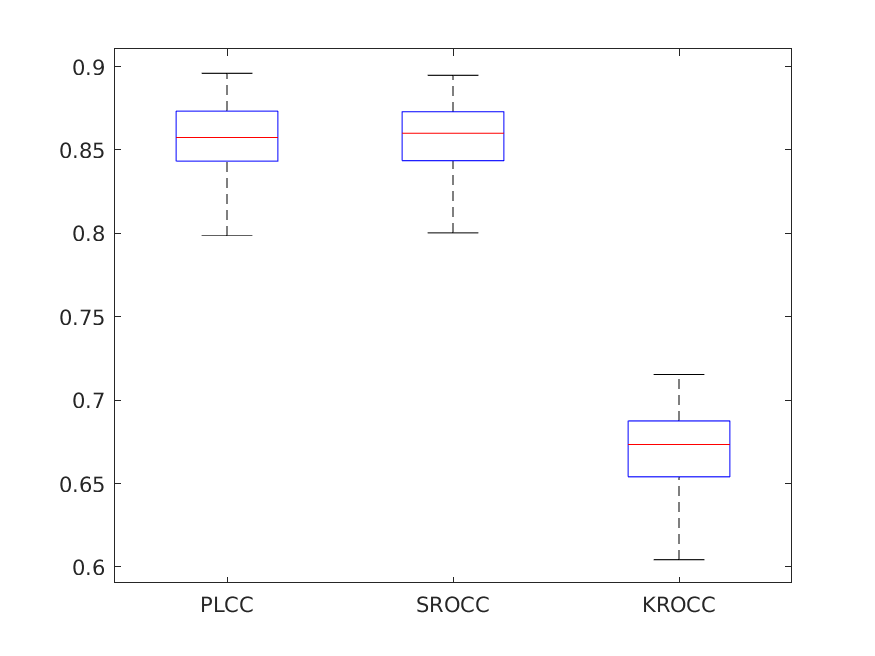}
        \caption{SCORER.}
    \end{subfigure}
    ~
    \begin{subfigure}[b]{0.45\textwidth}
        \includegraphics[width=\textwidth]{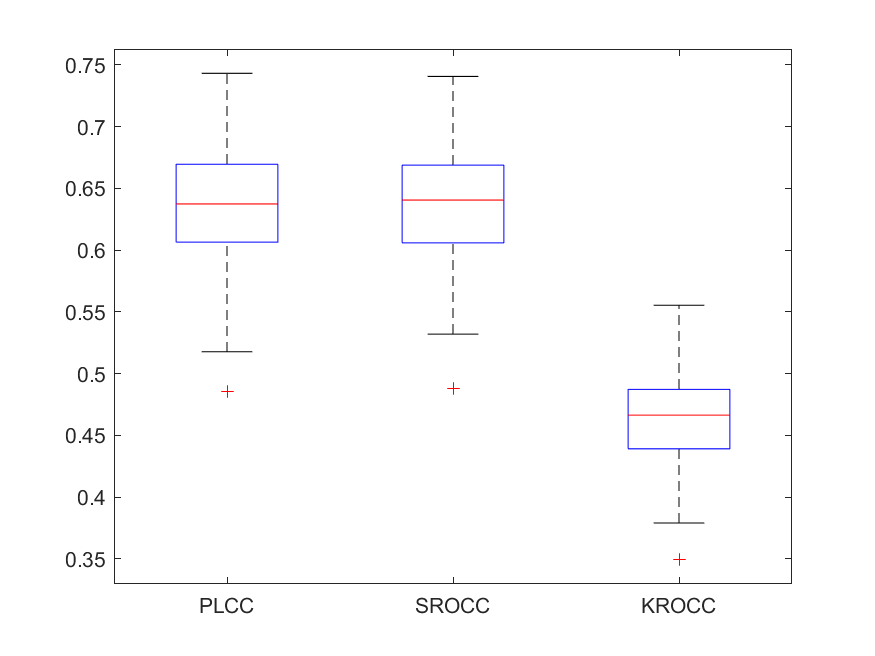}
        \caption{ENIQA.}
    \end{subfigure}
    
     \quad
    
    \begin{subfigure}[b]{0.45\textwidth}
        \includegraphics[width=\textwidth]{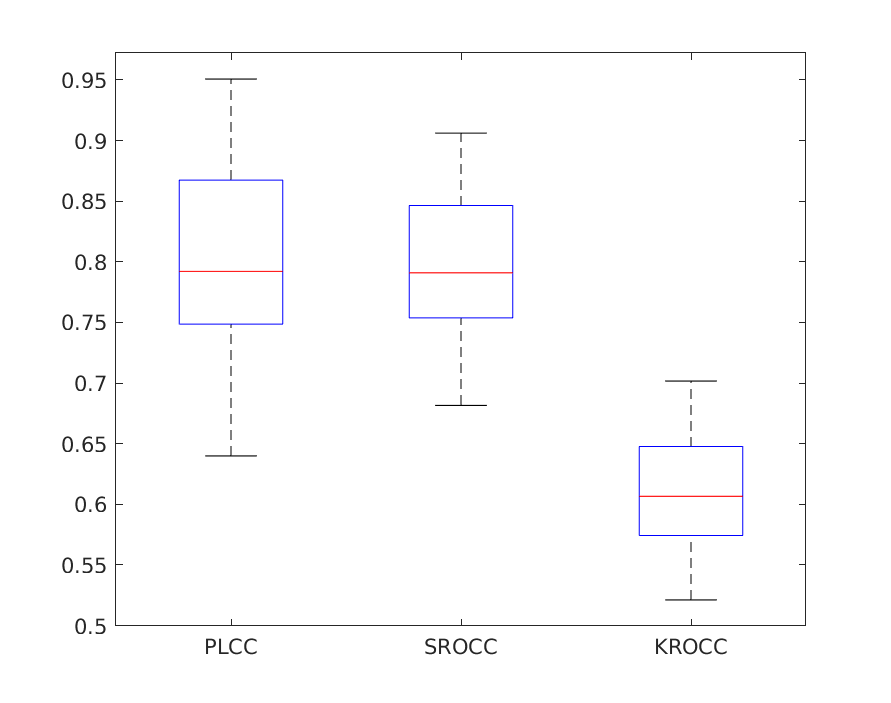}
        \caption{MultiGAP-SVR.}
    \end{subfigure}
    ~
    \begin{subfigure}[b]{0.45\textwidth}
        \includegraphics[width=\textwidth]{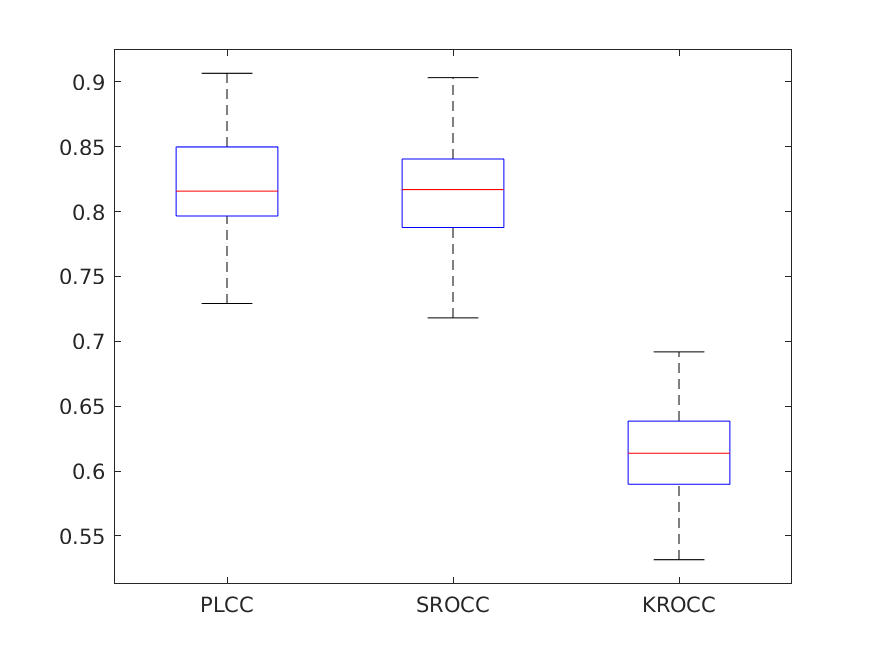}
        \caption{MultiGAP-GPR.}
    \end{subfigure}
    
    \caption{Box plots of the measured PLCC, SROCC, and KROCC values. 100 random train-test splits.}
    \label{fig:boxplot2}
\end{figure}

The performance of the examined algorithms in terms of average PLCC, SROCC, and KROCC
(100 random train-test splits) is summarized in Table \ref{table:overall}.
From these results, it can be observed that SCORER \cite{oszust2019local}
performs significantly
better than any other NR-IQA algorithms. MultiGAP-GPR \cite{varga2020multi} took
the second, MultiGAP-SVR \cite{varga2020multi} took the third,
and SPF-IQA \cite{varga2020no}
took the fourth place, respectively. The boxplot figures of the 100 random train-test splits
are depicted in Figures \ref{fig:boxplot1} and \ref{fig:boxplot2}. On each box, the central
mark indicates the median, and the bottom and top edges of the box indicate the 25$th$ and
75$th$ percentiles, respectively. The whiskers extend to the most extreme data points not
considered outliers. Moreover, the outliers are plotted by '+'.

\begin{table*}[ht]
\caption{Overall performance on KADID-10k \cite{lin2019kadid}. Average PLCC, SROCC, and KROCC
are reported, measured 100 random train-test splits. The best results are typed
by \textbf{bold}, the second best results are typed by \textit{italic}.
} 
\centering 
\begin{center}
    \begin{tabular}{ |c|c|c|c|}
    \hline
Method&PLCC&SROCC&KROCC\\
    \hline
DIIVINE \cite{moorthy2011blind}   &0.423 &0.428 &0.302 \\
BLIINDS-II \cite{saad2012blind} &0.548 &0.530 &0.377 \\
BRISQUE \cite{mittal2012no}   & 0.383& 0.386& 0.269\\
NIQE \cite{mittal2012making}  &0.273 &0.309 &0.309 \\
CurveletQA \cite{liu2014no} & 0.473& 0.450& 0.318\\
SSEQ \cite{liu2014noS}      &0.453 &0.433 &0.302 \\
GRAD-LOG-CP \cite{xue2014blind} &0.585 &0.566 &0.411 \\
PIQE \cite{venkatanath2015blind}      &0.289 &0.237 &0.237 \\
IL-NIQE \cite{zhang2015feature}  & 0.230& 0.211& 0.211\\
BMPRI \cite{min2018blind} & 0.554& 0.530& 0.379\\
SPF-IQA \cite{varga2020no} & 0.717& 0.708& 0.526\\
SCORER \cite{oszust2019local}    & \textbf{0.855}& \textbf{0.856}& \textbf{0.669}\\
ENIQA \cite{chen2019no} & 0.634& 0.636& 0.464\\
MultiGAP-SVR \cite{varga2020multi} &0.799 &0.795 &0.608 \\
MultiGAP-GPR \cite{varga2020multi} &\textit{0.820} &\textit{0.814} &\textit{0.613} \\
 \hline
 \end{tabular}
\end{center}
\label{table:overall}
\end{table*}

\subsection{Performance over different distortion levels}
\label{sec:level}
In this subsection, the performance of the examined NR-IQA algorithms
is given over different distortion levels.
Specifically, Tables \ref{table:levelPLCC},
\ref{table:levelSROCC}, and \ref{table:levelKROCC} summarize
the average PLCC, SROCC, and KROCC values over the different distortion
levels of KADID-10k database \cite{lin2019kadid}.
\begin{table*}[ht]
\caption{Performance over different distortion levels of KADID-10k \cite{lin2019kadid}.
Average PLCC is reported, measured over 100 random train-test splits.
} 
\centering 
\begin{center}
    \begin{tabular}{ |c|c|c|c|c|c|}
    \hline
&Level1&Level2&Level3&Level4&Level5\\
    \hline
DIIVINE \cite{moorthy2011blind}& 0.026&0.158&0.281&0.365&0.455 \\
BLIINDS-II \cite{saad2012blind}& 0.163&0.239&0.376&0.503&0.558\\
BRISQUE \cite{mittal2012no}& 0.071&0.148&0.198&0.339&0.441\\
NIQE \cite{mittal2012making}& 0.036&0.061&0.045&0.101&0.137\\
CurveletQA \cite{liu2014no}&0.088&0.185&0.317&0.412&0.495\\
SSEQ \cite{liu2014noS}&0.055&0.143&0.265&0.360&0.498\\
GRAD-LOG-CP \cite{xue2014blind}& 0.126&0.300&0.412&0.495&0.562\\
PIQE \cite{venkatanath2015blind}& 0.032&-0.007&0.048&0.115&0.248\\
IL-NIQE \cite{zhang2015feature}&0.003&0.084&0.127&0.153&0.120\\
BMPRI \cite{min2018blind}&0.097&0.265&0.399&0.478&0.545\\
SPF-IQA \cite{varga2020no}& 0.241&0.465&0.606&0.694&0.721\\
SCORER \cite{oszust2019local}& 0.490&0.742&0.806&0.844&0.787\\
ENIQA \cite{chen2019no}& 0.165&0.382&0.504&0.595&0.644\\
MultiGAP-SVR \cite{varga2020multi}&0.336 &0.472 &0.631 &0.725 &0.782\\
MultiGAP-GPR \cite{varga2020multi}&0.424 &0.577 &0.730 &0.825 &0.889\\
 \hline
 \end{tabular}
\end{center}
\label{table:levelPLCC}
\end{table*}

\begin{table*}[ht]
\caption{Performance over different distortion levels of KADID-10k \cite{lin2019kadid}.
Average SROCC is reported, measured over 100 random train-test splits.
} 
\centering 
\begin{center}
    \begin{tabular}{ |c|c|c|c|c|c|}
    \hline
&Level1&Level2&Level3&Level4&Level5\\
    \hline
DIIVINE \cite{moorthy2011blind}& -0.007&0.152&0.285&0.411&0.528 \\
BLIINDS-II \cite{saad2012blind}&0.172&0.228&0.358&0.535&0.629\\
BRISQUE \cite{mittal2012no}& 0.064&0.150&0.199&0.360&0.473\\
NIQE \cite{mittal2012making}& 0.141&0.126&0.065&0.104&0.162\\
CurveletQA \cite{liu2014no}&0.082&0.186&0.309&0.417&0.532\\
SSEQ \cite{liu2014noS}& 0.007&0.127&0.246&0.363&0.548\\
GRAD-LOG-CP \cite{xue2014blind}& 0.103&0.298&0.403&0.513&0.605\\
PIQE \cite{venkatanath2015blind}& 0.052&-0.005&0.007&0.061&0.207\\
IL-NIQE \cite{zhang2015feature}& 0.024&0.095&0.126&0.129&0.100\\
BMPRI \cite{min2018blind}&0.096&0.256&0.386&0.495&0.588\\
SPF-IQA \cite{varga2020no}&0.212&0.458&0.603&0.691&0.741\\
SCORER \cite{oszust2019local}& 0.436&0.737&0.818&0.839&0.769\\
ENIQA \cite{chen2019no}&0.127&0.373&0.505&0.610&0.688\\
MultiGAP-SVR \cite{varga2020multi}&0.331 &0.475 &0.662 &0.781 &0.842\\
MultiGAP-GPR \cite{varga2020multi}&0.407 &0.563 & 0.728&0.839 &0.906\\
 \hline
 \end{tabular}
\end{center}
\label{table:levelSROCC}
\end{table*}

\begin{table*}[ht]
\caption{Performance over different distortion levels of KADID-10k \cite{lin2019kadid}.
Average KROCC is reported, measured over 100 random train-test splits.
} 
\centering 
\begin{center}
    \begin{tabular}{ |c|c|c|c|c|c|}
    \hline
&Level1&Level2&Level3&Level4&Level5\\
    \hline
DIIVINE \cite{moorthy2011blind}&0.026&0.158&0.281&0.365&0.455 \\
BLIINDS-II \cite{saad2012blind}& 0.163&0.239&0.376&0.503&0.558\\
BRISQUE \cite{mittal2012no}&0.071&0.148&0.198&0.339&0.441\\
NIQE \cite{mittal2012making}& 0.036&0.061&0.045&0.101&0.137\\
CurveletQA \cite{liu2014no}&0.088&0.185&0.317&0.412&0.495\\
SSEQ \cite{liu2014noS}& 0.055&0.143&0.265&0.360&0.498\\
GRAD-LOG-CP \cite{xue2014blind}&0.126&0.300&0.412&0.495&0.562\\
PIQE \cite{venkatanath2015blind}& 0.032&-0.007&0.048&0.115&0.248\\
IL-NIQE \cite{zhang2015feature}&0.003&0.084&0.127&0.153&0.120\\
BMPRI \cite{min2018blind}& 0.097&0.265&0.399&0.478&0.545\\
SPF-IQA \cite{varga2020no}&0.241&0.465&0.606&0.694&0.721\\
SCORER \cite{oszust2019local}&0.490&0.742&0.806&0.844&0.787\\
ENIQA \cite{chen2019no}& 0.165&0.382&0.504&0.595&0.644\\
MultiGAP-SVR \cite{varga2020multi}&0.282 &0.382 &0.516 &0.605 &0.651\\
MultiGAP-GPR \cite{varga2020multi}&0.332 &0.442 &0.565 &0.653 &0.709\\
 \hline
 \end{tabular}
\end{center}
\label{table:levelKROCC}
\end{table*}

\subsection{Performance over different distortion types}
\label{sec:distortion}
In this subsection, the performance of the examined NR-IQA algorithms is
given over different distortion types. Specifically, Tables
\ref{table:distPLCC}, \ref{table:distSROCC}, and \ref{table:distKROCC}
summarize the average PLCC, SROCC, and KROCC values over the different
distortion types of KADID-10k database \cite{lin2019kadid}.
As already mentioned, Table \ref{table:distortions} summarizes the
different distortions types found in KADID-10k \cite{lin2019kadid} and
their numeric codes.
\begin{sidewaystable*}[ht]
\tiny
\caption{Performance over different distortion types of KADID-10k \cite{lin2019kadid}.
Average PLCC is reported, measured over 100 random train-test splits.
} 
\centering 
\begin{center}
    \begin{tabular}{ |c|c|c|c|c|c|c|c|c|c|c|c|c|c|c|c|c|c|c|c|c|c|c|c|c|c|}
    \hline
&1&2&3&4&5&6&7&8&9&10&11&12&13&14&15&16&17&18&19&20&21&22&23&24&25\\
    \hline
DIIVINE \cite{moorthy2011blind}&0.691&0.818&0.581&0.249&0.300&0.433&-0.020&0.086&0.558&0.672&0.454&0.541&0.459&0.480&0.777&0.204&0.221&0.037&0.612&0.065&0.432&0.383&0.041&0.608&0.105\\
BLIINDS-II \cite{saad2012blind}&0.799&0.772&0.412&0.602&0.016&0.499&0.094&0.530&0.690&0.806&0.529&0.684&0.583&0.563&0.685&0.438&0.556&0.232&0.843&0.045&0.651&0.315&0.166&0.583&0.116\\
BRISQUE \cite{mittal2012no}&0.764&0.808&0.461&0.192&0.180&0.403&0.091&0.240&0.057&0.598&0.350&0.437&0.411&0.428&0.642&0.433&0.258&0.093&0.679&0.135&0.416&0.378&0.098&0.507&0.039\\
NIQE \cite{mittal2012making}&0.760&0.291&0.628&0.242&-0.123&0.585&-0.044&0.244&0.647&0.930&0.641&0.746&0.791&0.655&0.807&0.516&0.538&0.242&0.869&0.102&0.466&0.759&0.056&0.440&0.180\\
CurveletQA \cite{liu2014no}&0.821&0.851&0.714&0.228&0.109&0.653&0.020&0.081&0.665&0.634&0.691&0.719&0.591&0.612&0.777&0.426&0.324&0.078&0.634&0.078&0.142&0.339&0.037&0.589&0.027\\
SSEQ \cite{liu2014noS}&0.716&0.723&0.368&0.366&0.031&0.556&0.075&0.210&0.560&0.800&0.622&0.664&0.556&0.581&0.570&0.315&0.395&0.115&0.571&0.026&0.523&0.218&0.165&0.566&0.086\\
GRAD-LOG-CP \cite{xue2014blind}&0.826&0.842&0.526&0.363&0.021&0.735&-0.028&0.320&0.764&0.887&0.816&0.844&0.685&0.723&0.845&0.470&0.463&0.168&0.839&0.072&0.778&0.598&0.302&0.674&0.138\\
PIQE \cite{venkatanath2015blind}&0.902&0.773&0.311&0.264&-0.052&0.700&0.067&0.164&0.813&0.820&0.774&0.865&0.601&0.815&-0.234&0.468&0.501&0.117&-0.293&0.047&0.403&0.696&-0.016&0.376&-0.059\\
IL-NIQE \cite{zhang2015feature}&0.475&0.488&0.277&-0.092&0.087&0.160&0.115&0.028&0.230&0.382&0.508&0.666&0.518&0.477&0.676&-0.128&-0.200&-0.035&0.378&-0.014&0.324&-0.295&0.015&0.243&0.045\\
BMPRI \cite{min2018blind}&0.844&0.810&0.393&0.386&0.117&0.704&0.094&0.431&0.723&0.927&0.807&0.729&0.424&0.565&0.832&0.478&0.487&0.241&0.718&-0.016&0.677&0.334&0.141&0.495&0.130\\
SPF-IQA \cite{varga2020no}&0.845&0.829&0.546&0.800&0.289&0.761&0.124&0.761&0.685&0.865&0.880&0.901&0.699&0.750&0.884&0.686&0.502&0.143&0.726&0.080&0.791&0.704&0.354&0.809&0.236\\
SCORER \cite{oszust2019local}&0.925&0.929&0.846&0.813&0.722&0.776&0.030&0.820&0.896&0.929&0.897&0.928&0.894&0.902&0.927&0.773&0.784&0.267&0.913&0.318&0.787&0.798&0.483&0.806&0.362\\
ENIQA \cite{chen2019no}&0.778&0.792&0.576&0.719&0.226&0.651&0.035&0.687&0.709&0.847&0.748&0.769&0.608&0.708&0.782&0.570&0.470&0.159&0.668&0.011&0.546&0.537&0.102&0.677&0.178\\
MultiGAP-SVR \cite{varga2020multi}&0.842&0.830&0.715&0.874&0.631&0.416&0.390&0.765&0.751&0.898&0.572&0.616&0.611&0.660&0.830&0.677&0.493&0.219&0.790&0.329&0.693&0.502&0.530&0.709&0.153\\
MultiGAP-GPR \cite{varga2020multi}&0.966&0.940&0.857&0.979&0.755&0.664&0.446&0.848&0.916&0.932&0.710&0.756&0.769&0.804&0.932&0.799&0.623&0.295&0.937&0.426&0.793&0.674&0.584&0.834&0.232\\
 \hline
 \end{tabular}
\end{center}
\label{table:distPLCC}
\end{sidewaystable*}

\begin{sidewaystable*}[ht]
\tiny
\caption{Performance over different distortion types of KADID-10k \cite{lin2019kadid}.
Average SROCC is reported, measured over 100 random train-test splits.
} 
\centering 
\begin{center}
    \begin{tabular}{ |c|c|c|c|c|c|c|c|c|c|c|c|c|c|c|c|c|c|c|c|c|c|c|c|c|c|}
    \hline
&1&2&3&4&5&6&7&8&9&10&11&12&13&14&15&16&17&18&19&20&21&22&23&24&25\\
    \hline
DIIVINE \cite{moorthy2011blind}&0.719&0.803&0.584&0.324&0.217&0.399&-0.008&0.089&0.515&0.624&0.473&0.543&0.453&0.504&0.801&0.176&0.166&0.026&0.597&0.067&0.394&0.386&0.051&0.636&0.072\\
BLIINDS-II \cite{saad2012blind}&0.791&0.761&0.405&0.529&0.029&0.475&0.099&0.504&0.639&0.763&0.550&0.688&0.612&0.580&0.699&0.405&0.422&0.212&0.818&0.032&0.576&0.313&0.161&0.616&0.112\\
BRISQUE \cite{mittal2012no}&0.781&0.791&0.449&0.255&0.142&0.397&0.087&0.251&0.051&0.556&0.396&0.457&0.382&0.482&0.692&0.424&0.196&0.105&0.665&0.116&0.376&0.363&0.107&0.418&0.048\\
NIQE \cite{mittal2012making}&0.743&0.255&0.652&0.313&-0.175&0.614&-0.041&0.235&0.704&0.850&0.648&0.755&0.854&0.698&0.870&0.445&0.411&0.179&0.871&0.111&0.667&0.769&0.050&0.731&0.148\\
CurveletQA \cite{liu2014no}&0.810&0.853&0.717&0.285&0.138&0.631&0.026&0.068&0.594&0.609&0.716&0.749&0.621&0.637&0.779&0.412&0.209&0.068&0.590&0.048&0.130&0.342&0.086&0.627&0.008\\
SSEQ \cite{liu2014noS}&0.716&0.731&0.368&0.436&0.026&0.538&0.063&0.206&0.462&0.695&0.641&0.678&0.597&0.606&0.607&0.259&0.303&0.102&0.554&0.004&0.482&0.206&0.174&0.593&0.061\\
GRAD-LOG-CP \cite{xue2014blind}&0.812&0.811&0.512&0.385&0.067&0.682&-0.002&0.323&0.677&0.788&0.847&0.863&0.704&0.741&0.826&0.437&0.372&0.143&0.796&0.083&0.687&0.590&0.305&0.698&0.132\\
PIQE \cite{venkatanath2015blind}&0.863&0.770&0.299&0.317&-0.008&0.727&0.064&0.159&0.758&0.768&0.811&0.875&0.582&0.828&-0.099&0.421&0.325&0.112&-0.271&0.009&0.353&0.717&-0.010&0.406&-0.028\\
IL-NIQE \cite{zhang2015feature}&0.519&0.596&0.275&-0.097&0.018&0.162&0.113&0.009&0.189&0.335&0.528&0.680&0.576&0.488&0.719&-0.134&-0.166&-0.028&0.363&-0.032&0.297&-0.303&0.035&0.247&0.026\\
BMPRI \cite{min2018blind}&0.843&0.817&0.394&0.453&0.106&0.669&0.118&0.444&0.618&0.820&0.837&0.753&0.461&0.598&0.823&0.438&0.374&0.197&0.699&0.004&0.540&0.296&0.146&0.520&0.133\\
SPF-IQA \cite{varga2020no}&0.835&0.802&0.537&0.791&0.306&0.722&0.110&0.770&0.601&0.802&0.889&0.914&0.714&0.773&0.883&0.644&0.382&0.140&0.709&0.096&0.715&0.703&0.369&0.821&0.225\\
SCORER \cite{oszust2019local}&0.898&0.885&0.841&0.751&0.661&0.761&0.053&0.811&0.844&0.812&0.902&0.923&0.892&0.911&0.916&0.732&0.658&0.230&0.874&0.332&0.701&0.778&0.408&0.810&0.388\\
ENIQA \cite{chen2019no}&0.785&0.795&0.572&0.673&0.162&0.629&0.050&0.671&0.636&0.773&0.766&0.790&0.611&0.732&0.814&0.528&0.354&0.100&0.651&0.006&0.482&0.527&0.119&0.702&0.170\\
MultiGAP-SVR \cite{varga2020multi}&0.915&0.895&0.775&0.898&0.665&0.406&0.387&0.832&0.746&0.902&0.623&0.670&0.660&0.714&0.916&0.681&0.437&0.266&0.842&0.358&0.696&0.520&0.568&0.810&0.173\\
MultiGAP-GPR \cite{varga2020multi}&0.975&0.964&0.873&0.983&0.758&0.634&0.433&0.872&0.837&0.980&0.721&0.771&0.791&0.823&0.965&0.747&0.523&0.310&0.928&0.432&0.736&0.644&0.600&0.893&0.238\\
\hline
 \end{tabular}
\end{center}
\label{table:distSROCC}
\end{sidewaystable*}

\begin{sidewaystable*}[ht]
\tiny
\caption{Performance over different distortion types of KADID-10k \cite{lin2019kadid}.
Average KROCC is reported, measured over 100 random train-test splits.
} 
\centering 
\begin{center}
    \begin{tabular}{ |c|c|c|c|c|c|c|c|c|c|c|c|c|c|c|c|c|c|c|c|c|c|c|c|c|c|}
    \hline
&1&2&3&4&5&6&7&8&9&10&11&12&13&14&15&16&17&18&19&20&21&22&23&24&25\\
    \hline
DIIVINE \cite{moorthy2011blind}&0.691&0.818&0.581&0.249&0.300&0.433&-0.020&0.086&0.558&0.672&0.454&0.541&0.459&0.480&0.777&0.204&0.221&0.037&0.612&0.065&0.432&0.383&0.041&0.608&0.105\\
BLIINDS-II \cite{saad2012blind}&0.799&0.772&0.412&0.602&0.016&0.499&0.094&0.530&0.690&0.806&0.529&0.684&0.583&0.563&0.685&0.438&0.556&0.232&0.843&0.045&0.651&0.315&0.166&0.583&0.116\\
BRISQUE \cite{mittal2012no}&0.764&0.808&0.461&0.192&0.180&0.403&0.091&0.240&0.057&0.598&0.350&0.437&0.411&0.428&0.642&0.433&0.258&0.093&0.679&0.135&0.416&0.378&0.098&0.507&0.039\\
NIQE \cite{mittal2012making}&0.760&0.291&0.628&0.242&-0.123&0.585&-0.044&0.244&0.647&0.930&0.641&0.746&0.791&0.655&0.807&0.516&0.538&0.242&0.869&0.102&0.466&0.759&0.056&0.440&0.180\\
CurveletQA \cite{liu2014no}&0.821&0.851&0.714&0.228&0.109&0.653&0.020&0.081&0.665&0.634&0.691&0.719&0.591&0.612&0.777&0.426&0.324&0.078&0.634&0.078&0.142&0.339&0.037&0.589&0.027\\
SSEQ \cite{liu2014noS}&0.716&0.723&0.368&0.366&0.031&0.556&0.075&0.210&0.560&0.800&0.622&0.664&0.556&0.581&0.570&0.315&0.395&0.115&0.571&0.026&0.523&0.218&0.165&0.566&0.086\\
GRAD-LOG-CP \cite{xue2014blind}&0.826&0.842&0.526&0.363&0.021&0.735&-0.028&0.320&0.764&0.887&0.816&0.844&0.685&0.723&0.845&0.470&0.463&0.168&0.839&0.072&0.778&0.598&0.302&0.674&0.138\\
PIQE \cite{venkatanath2015blind}&0.902&0.773&0.311&0.264&-0.052&0.700&0.067&0.164&0.813&0.820&0.774&0.865&0.601&0.815&-0.234&0.468&0.501&0.117&-0.293&0.047&0.403&0.696&-0.016&0.376&-0.059\\
IL-NIQE \cite{zhang2015feature}&0.475&0.488&0.277&-0.092&0.087&0.160&0.115&0.028&0.230&0.382&0.508&0.666&0.518&0.477&0.676&-0.128&-0.200&-0.035&0.378&-0.014&0.324&-0.295&0.015&0.243&0.045\\
BMPRI \cite{min2018blind}&0.844&0.810&0.393&0.386&0.117&0.704&0.094&0.431&0.723&0.927&0.807&0.729&0.424&0.565&0.832&0.478&0.487&0.241&0.718&-0.016&0.677&0.334&0.141&0.495&0.130\\
SPF-IQA \cite{varga2020no}&0.845&0.829&0.546&0.800&0.289&0.761&0.124&0.761&0.685&0.865&0.880&0.901&0.699&0.750&0.884&0.686&0.502&0.143&0.726&0.080&0.791&0.704&0.354&0.809&0.236\\
SCORER \cite{oszust2019local}&0.925&0.929&0.846&0.813&0.722&0.776&0.030&0.820&0.896&0.929&0.897&0.928&0.894&0.902&0.927&0.773&0.784&0.267&0.913&0.318&0.787&0.798&0.483&0.806&0.362\\
ENIQA \cite{chen2019no}&0.778&0.792&0.576&0.719&0.226&0.651&0.035&0.687&0.709&0.847&0.748&0.769&0.608&0.708&0.782&0.570&0.470&0.159&0.668&0.011&0.546&0.537&0.102&0.677&0.178\\
MultiGAP-SVR \cite{varga2020multi}&0.716&0.704&0.600&0.702&0.523&0.334&0.318&0.642&0.582&0.711&0.489&0.524&0.518&0.560&0.722&0.534&0.354&0.233&0.651&0.298&0.547&0.418&0.441&0.632&0.173\\
MultiGAP-GPR \cite{varga2020multi}&0.778&0.765&0.685&0.796&0.598&0.501&0.353&0.684&0.664&0.793&0.572&0.605&0.622&0.649&0.773&0.589&0.418&0.268&0.730&0.355&0.580&0.516&0.469&0.710&0.220\\
 \hline
 \end{tabular}
\end{center}
\label{table:distKROCC}
\end{sidewaystable*}

\section{Conclusion}
\label{sec:conclusion}
In this study, several NR-IQA algorithms, including DIIVINE \cite{moorthy2011blind},
BLIINDS-II \cite{saad2012blind}, BRISQUE \cite{mittal2012no},
NIQE \cite{mittal2012making}, CurveletQA \cite{liu2014no}, SSEQ \cite{liu2014noS},
GRAD-LOG-CP \cite{xue2014blind}, PIQE \cite{venkatanath2015blind},
IL-NIQE \cite{zhang2015feature}, BMPRI \cite{min2018blind},
SPF-IQA \cite{varga2020no}, SCORER \cite{oszust2019local},
ENIQA \cite{chen2019no}, MultiGAP-SVR \cite{varga2020multi}, and
MultiGAP-GPR \cite{varga2020multi}, were evaluated on KADID-10k \cite{lin2019kadid}
database.
\bibliographystyle{unsrt}  

\bibliography{references}

\end{document}